\documentstyle[psfig]{mn}

\newif\ifAMStwofonts

\newcommand{\lapp}{\mbox{\raisebox{-0.3em}{$\stackrel{\textstyle <}{\sim}$}}}
\newcommand{\gapp}{\mbox{\raisebox{-0.3em}{$\stackrel{\textstyle >}{\sim}$}}}

\title{J0041+3224: a new double-double radio galaxy}
\author[D.J. Saikia et al.]
       {D.J. Saikia$\thanks{E-mail: djs@ncra.tifr.res.in}$, C. Konar 
        and V.K. Kulkarni \\
National Centre for Radio Astrophysics, TIFR, Pune University Campus, Post Bag 3, Pune 411 007, India}
\date{Accepted.    Received }

\pagerange{\pageref{firstpage}--\pageref{lastpage}}
\pubyear{  }

\begin{document}

\maketitle

\label{firstpage}

\begin{abstract}
We report the discovery of a double-double radio galaxy (DDRG), J0041+3224, with the 
Giant Metrewave Radio Telescope (GMRT) and subsequent high-frequency observations with
the Very Large Array (VLA). The inner and outer doubles are aligned within $\sim$4$^\circ$
and are reasonably collinear with the parent optical galaxy. The outer double has a steeper radio 
spectrum compared with the inner one. Using an estimated redshift of 0.45, the projected
linear sizes of the outer and inner doubles are 969 and 171 kpc respectively. The time scale 
of interruption of jet activity has been estimated to be $\sim$20 Myr, similar to
other known DDRGs.  We have compiled a sample of known DDRGs, and have re-examined the 
inverse correlation between the ratio of the luminosities of the outer to the inner double
and the size of the inner double, $l_{in}$. 
Unlike the other DDRGs with $l_{in}$$\gapp$50 kpc, the inner double of J0041+3224 
is marginally more luminous than the outer one. The two DDRGs with  $l_{in}$$\lapp$few kpc 
have a more luminous inner double than the outer one, possibly due to a higher efficiency of 
conversion of beam energy as the jets propagate through the dense interstellar medium. 
We have examined the symmetry parameters and find that the inner doubles 
appear to be more asymmetric in both its armlength and flux density ratios compared
with the outer doubles, although they appear marginally more collinear with the core 
than the outer double. We discuss briefly possible implications of these trends. 
\end{abstract}


\section{Introduction}
One of the important issues concerning galaxies is the duration of their active 
galactic nuclei (AGN) phase and whether such periods of activity are
episodic. In the currently widely accepted paradigm, activity is believed
to be intimately related to the `feeding' of a supermassive black hole whose
mass ranges from $\sim$10$^6$ to 10$^{10}$ M$_\odot$. Such an active phase 
may be recurrent with  an average total timescale of the active phases being
$\sim$10$^8$ to 10$^{9}$ yr (cf. Marconi et al. 2004, and references therein). 

Of the galaxies harbouring an AGN, a small fraction 
appears to be luminous at radio wavelengths. For example, in the SDSS
(Sloan Digital Sky Survey) quasars, $\sim$8 per cent of the bright ones ($i<$18.5)
are radio loud in the sense that the ratio of radio to X-ray flux exceeds
unity (Ivezi\'{c} et al. 2002, 2004). Although what physical conditions determine 
loudness still remains unclear, Nipoti, Blundell \& Binney (2005) have 
suggested recently that this may simply be a function of the epoch at which 
the source is observed. 

For the radio-loud objects, an interesting way of probing their history 
is via the structural and spectral information of the lobes of extended 
radio emission. Such studies have been used to probe sources which exhibit 
precession or changes in the ejection axis, effects of motion of the parent galaxy, 
backflows from hotspots as well as X-shaped sources and major interruptions of jet 
activity. For example, the radio galaxy 3C388 exhibits two distinct regions
of emission separated by a jump in spectral index, which has been interpreted
to be due to two different epochs of jet activity (Burns, Schwendeman \& White 1983;
Roettiger et al. 1994). A ridge of emission, reminescent of a jet but displaced
towards the south of the nucleus in the radio galaxy 3C338 could also be due
to intermittent jet activity (Burns, Schwendeman \& White 1983). Other suggestions
of distinct epochs of jet activity based on spectral index studies include
Her A, where the bright inner regions have flatter spectra with a sharp boundary
delineating it from the more extended lobe emission, especially in the western lobe 
(Gizani \& Leahy 2003), and 3C310 where the inner
components B and D have substanially flatter spectra than the surrounding lobes
(van Breugel \& Fomalont 1984; Leahy et al. 1986). An interesting example of
different epochs of jet activity is the well-studied radio galaxy Cen A, where in
addition to the diffuse outer lobes there are the more compact inner lobes and
a northern middle lobe or NML (Burns, Feigelson \& Schreier 1983; Clarke, Burns \& Norman
1992; Junkes et al. 1993). Morganti et al. (1999) have detected a large-scale jet
connecting the northern lobe of the inner double and the NML, and have suggested that 
the formation of the NML may be due to a `bursting bubble' in which plasma
accumulated in the inner lobe bursts out through a nozzle. A lobe of emission
on only one side of the nuclear region has also been seen in the Gigahertz Peaked
Spectrum radio source B0108+388, and has been suggested to be a relic of a previous
cycle of jet activity (Baum et al. 1990). However, the one-sidedness of the emission
is puzzling (cf. Stanghellini et al. 2005), and it would be interesting to examine
whether a `bursting bubble' model may also be applicable in such cases. 

One of the more
striking examples of episodic jet activity is when a new pair of radio lobes
is seen closer to the nucleus before the `old' and more distant radio
lobes have faded. Such sources have been christened as `double-double' radio galaxies
(DDRGs) by Schoenmakers et al. (2000, hereinafter referred to as S2000). 
They proposed a relatively general definition of a DDRG as a double-double radio
galaxy consisting of a pair of double radio sources with a common centre. S2000 also
suggested that the two lobes of the inner double should have an edge-brightened
radio morphology to distinguish it from knots in a jet. In such sources the newly-formed
jets propagate outwards through the cocoon formed by the earlier cycle of activity rather 
than the general intergalactic or intracluster medium, after traversing through the 
interstellar medium of the host galaxy. Approximately a dozen or so of such DDRGs are known
in the literature (S2000; Saripalli et al. 2002, 2003; Schilizzi et al. 2001;
Marecki et al. 2003). We have included objects such as 3C236 
(Schilizzi et al. 2001) and J1247+6723 (Marecki et al. 2003) in this category since the 
principal difference is only in the size of the inner double which is $\lapp$few kpc. It is 
important to identify more DDRGs not only for understanding episodic jet activity and examining 
their time scales, but also for studying the 
propagation of jets in different media. For example, to explain the edge-brightened 
hotspots in the inner doubles of the DDRGs, Kaiser, Schoenmakers \& R\"{o}ttgering (2000) 
have suggested that warm (T$\sim$10$^4$K) clouds of gas in the intergalactic medium are 
dispersed over the cocoon volume by surface instabilities induced by the passage of
the cocoon material.

In this paper we report the discovery of a new DDRG, J0041+3224, identified from 
observations made with the Giant Metrewave Radio Telescope (GMRT) of candidate
DDRGs from the B2 sample (Padrielli, Kapahi \& Katgert-Merkelijn 1981; Saikia et al. 2002). 
Our candidates were identified by comparing the large- and smaller-scale images made either by us
or those available in the literature. J0041+3224 was reported by Padrielli et al. (1981)
to be double-lobed with an angular size of 32 arcsec.  They identified the radio source 
to be associated with a galaxy with a visual magnitude of 20.0 and located at 
RA 00$^h$ 41$^m$ 46.$^s$11, Dec: +32$^\circ$ 24$^\prime$ 53.$^{\prime\prime}$8 in J2000 co-ordinates.
There is no measured redshift of the galaxy. From the V magnitude$-$redshift diagram
(Guiderdoni \& Rocca-Volmerange 1987) we estimate the redshift to be $\sim$0.45 and use
this value for this paper.

\section{Observations and analyses} 
The GMRT consists of thirty 45-m antennas in an approximate `Y'
shape similar to the VLA but with each antenna in a fixed position.
Twelve antennas are randomly placed within a central
1 km by 1 km square (the ``Central Square'') and the
remainder form the irregularly shaped Y (6 on each arm) over a total extent of about 25 km.
Further details about the array can be found at the GMRT website
at {\tt http://www.gmrt.ncra.tifr.res.in}.
The observations were made in the standard fashion, with
each source observation interspersed with observations
of the phase calibrator.  The flux densities are on the Baars et al. (1977) scale. 

The observations with the VLA were made in the snap-shot mode in the
L, C and X bands. The flux densities are again on the Baars et al. (1977) scale.  
All the data were calibrated and analysed in the standard way using the NRAO {\tt AIPS}
package.
                                                                                                                       
The observing log for both the GMRT and the VLA observations are listed in
Table 1 which is arranged as follows. Columns 1 and 2 show the name of the
telescope, and the array configuration for the VLA observations;
column 3 shows the frequency of the observations, while the 
dates of the observations are listed in column 4.

\begin{table}
\caption{ Observing log }
\begin{tabular}{l c r c }
                                                                                                                       
\hline
Teles-    & Array  & Obs.   &   Obs. Date  \\
cope      & Conf.  & Freq.  &              \\
          &        & MHz    &               \\
\hline
GMRT      &        & 617    & 2002 Jul 27   \\
GMRT      &        & 1287   & 2002 Jun 21    \\
VLA       &    C   & 1400   & 2002 Nov 28     \\
VLA       &    C   & 4860   & 2002 Nov 28     \\
VLA       &    C   & 8460   & 2002 Nov 28     \\
\hline
\end{tabular}
\end{table}

\begin{table*}
\caption{ The observational parameters and observed properties of the sources}
                                                                                                                          
\begin{tabular}{l rrr r r l rr l rr l rr l rr}
\hline
Freq.       & \multicolumn{3}{c}{Beam size}                    & rms      & S$_I$   & Cp  & S$_p$  & S$_t$  & Cp   & S$_p$ & S$_t$ & Cp  & S$_p$   & S$_t$   &  Cp  & S$_p$ & S$_t$   \\
                                                                                                                          
            MHz         & $^{\prime\prime}$ & $^{\prime\prime}$ & $^\circ$ &    mJy   & mJy     &     & mJy    & mJy    &      & mJy   & mJy   &     & mJy     & mJy   &   & mJy & mJy    \\
                        &                   &                   &          &  /b      &         &     & /b     &        &      & /b    &       &     & /b      &        &   &  /b &       \\
(1) & (2) & (3) & (4) & (5) & (6) & (7) & (8) & (9) & (10) & (11) & (12) & (13) & (14) & (15) & (16) & (17) & (18 ) \\
\hline
  G617       &  6.5    &  4.8       & 165                       &   0.49   &  2211   & W1  &  98    & 542    & W2   & 274   & 314   & E2  & 567     & 691  & E1 & 135 & 598     \\
                                                                                                                          
  G1287      &  2.6    &   2.3      &  25                       &   0.20   &  1104   & W1  &  12    & 233    & W2   &  92   & 145   & E2  & 231     & 373  & E1 &  13 & 227     \\
                                                                                                                          
 V1400       & 45.0    & 45.0       &                           &   0.51   &   967   &     &        &        &      &   &       &     &         &      &    &     &        \\
                                                                                                                          
 V1400       & 14.7    & 12.6       &  11                       &   0.16   &   940   & W1  &  108   &  207   & W2   & 139   & 148   & E2  &  338    & 377  & E1 & 117 & 202    \\
                                                                                                                          
 V4860       &  4.0    &  3.7       &  13                       &   0.03   &  298    & W1  & 7.4    &  50    & W2   & 42   & 52    & E2  &  126    & 154  & E1 & 6.0 &  37    \\
                                                                                                                          
 V8460       &  2.4    &  2.3       &  19                       &   0.01   &  153    & W1  & 1.4    &  12    & W2   & 21   &  32   & E2  &  75    & 100  & E1 & 0.9 & 7.6    \\
                                                                                                                          
\hline
\end{tabular}
\end{table*}

\begin{figure}
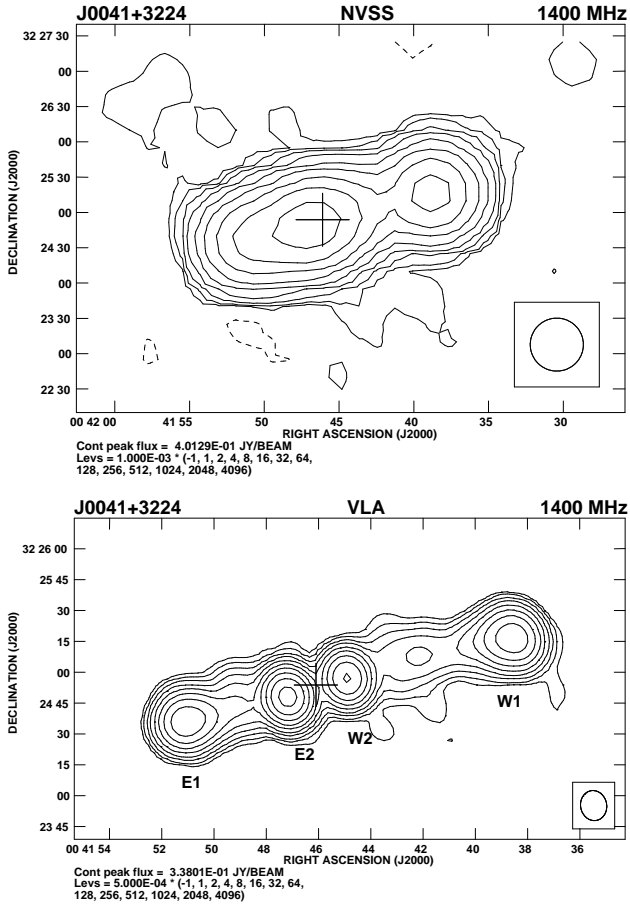

\vbox{
  \psfig{file=J0041_NVSS.ps,width=3.35in,angle=-90}
   \psfig{file=J0041_V1400_marked.ps,width=3.35in,angle=-90}
    }
\caption[]{The NVSS image of J0041+3224 with an angular resolution of 
45 arcsec (upper panel) and the VLA C-array image at 1400 MHz with an angular resolution of 
$\sim$13.6 arcsec. In all the images presented in this paper the restoring beam is indicated by an ellipse,
and the $+$ sign indicates the position of the optical galaxy.}
\end{figure}

\begin{figure}
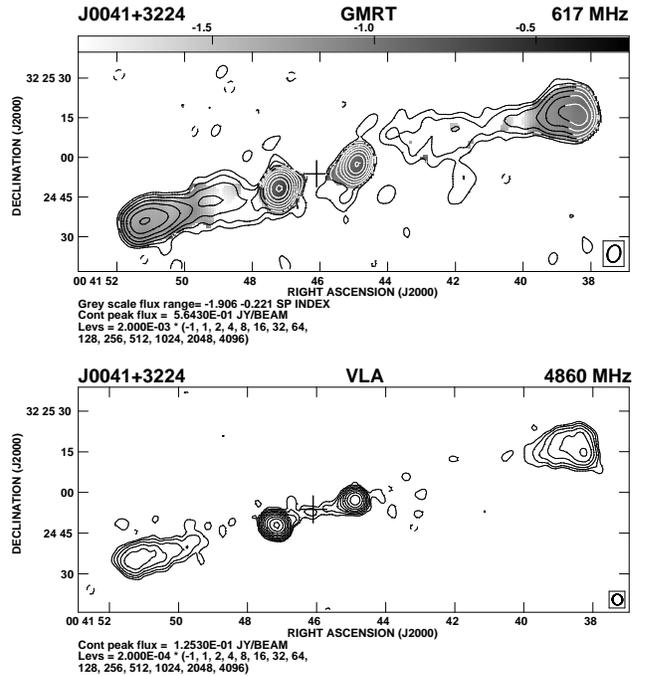

\vbox{
   \psfig{file=J0041_G610_spix.ps,width=3.35in,angle=-90}
   \psfig{file=J0041_VLA_CNEW.ps,width=3.35in,angle=-90}
    }
\caption[]{The GMRT image of J0041+3224 at 617 MHz with an angular resolution of
$\sim$5.6 arcsec (upper panel) and the VLA C-array image at 4860 MHz with an angular resolution of
$\sim$3.8 arcsec (lower panel). The spectral index image obtained by smoothing the 4860-MHz image to that
of the 617-MHz one is shown superimposed on the 617-MHz image in gray scale.}
\end{figure}

\begin{figure}
\vbox{
    \psfig{file=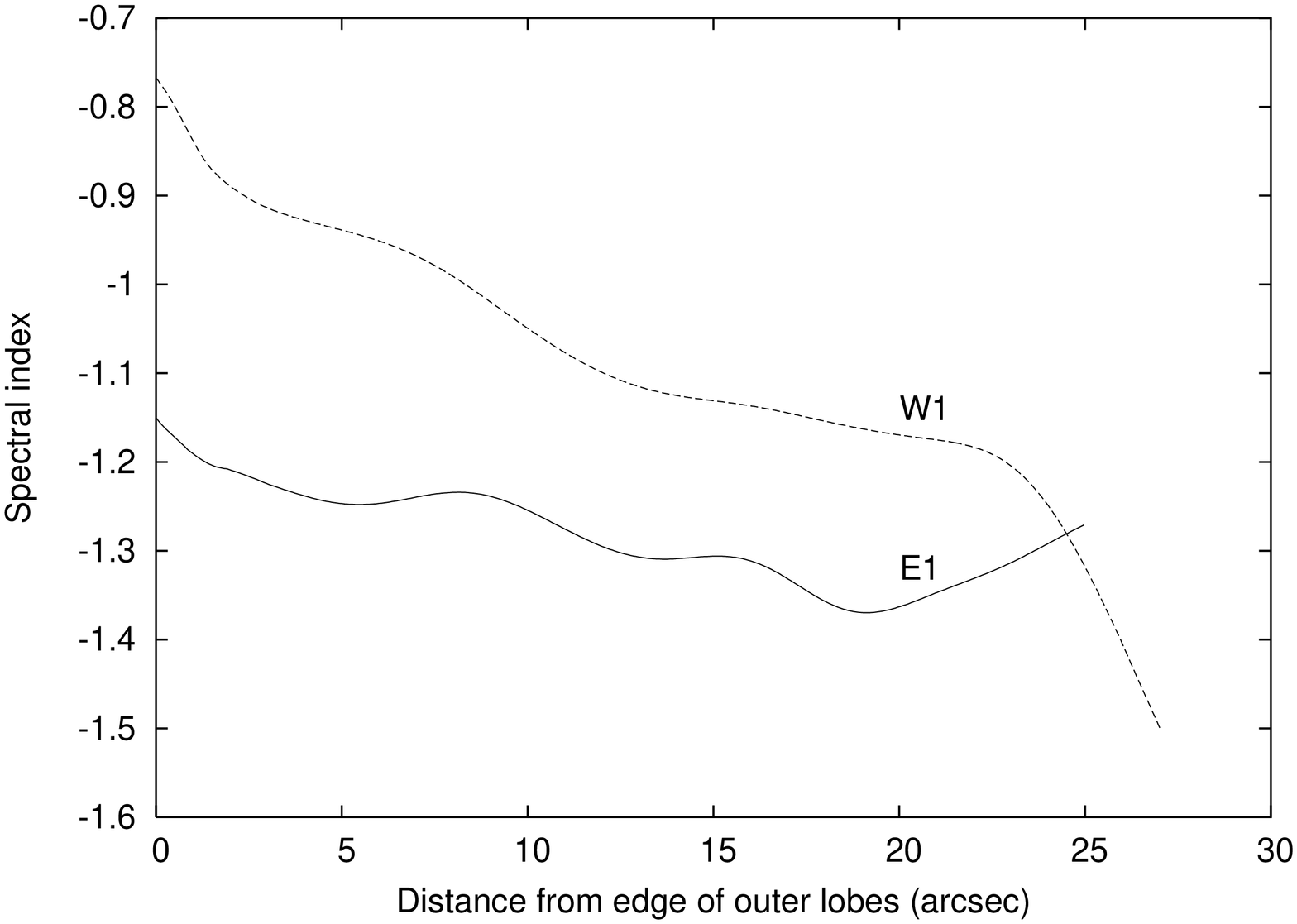,width=3.2in,angle=-90}
    }
\caption[]{Slices of spectral index between 617 and 4860 MHz for the western and eastern components of
the outer double.}
\end{figure}

\section{Observational results}
All the images of the source are presented in
Figs. 1, 2 and 4, while the observational parameters and some of the observed
properties are presented in Table 2, which is arranged as follows.
Column 1: frequency  of observations in units of MHz, with
the letter G or V representing either GMRT or VLA observations;
columns 2$-$4: the major and minor axes of the restoring beam in arcsec and its position 
angle (PA) in degrees;
column 5: the rms noise in units of mJy/beam; column 6: the integrated flux density of the
source in mJy estimated by specifying an area around the source; column 7, 10, 13 and 16: 
component designation where W1 and E1 indicate the western and eastern components of the outer 
double, W2 and E2 the western and eastern components of the inner double (Fig. 1, lower panel);  
columns 8 and 9, 11 and 12, 14 and 15, 17 and 18: the peak and total flux densities of the
components in units of mJy/beam and mJy respectively. The flux densities have been estimated
by specifying an area around each component. 
                                                                                                                     
The low-resolution NVSS image with an angular resolution of 45 arcsec (Fig. 1, upper panel) 
shows a more compact western component and a well-resolved eastern one.  The VLA C-array 
image at 1400 MHz (Fig. 1, lower panel) with an angular resolution of $\sim$13.6 arcsec 
shows the source could be a DDRG with the eastern NVSS component being 
resolved into the inner double and the eastern component of the outer double. 
The optical galaxy lies between the components of the inner double-lobed source. The PAs 
of the outer and inner doubles are 104$^\circ$ and 108$^\circ$ respectively, showing 
the close alignment seen in a number of other DDRGs (cf. S2000).  

\begin{figure}
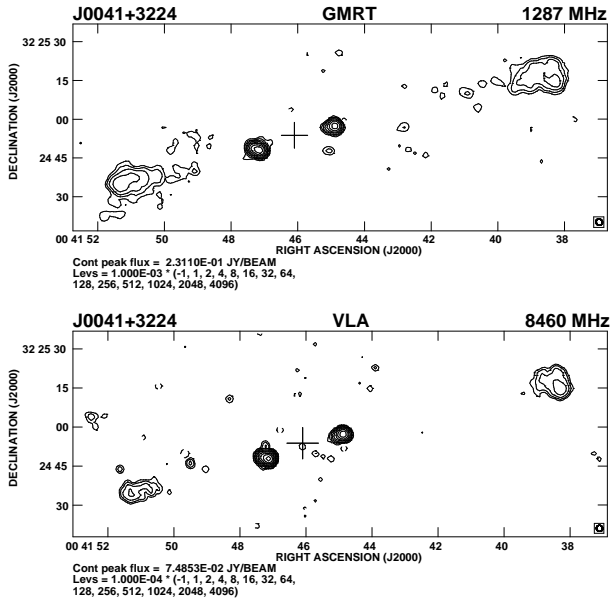

\vbox{
    \psfig{file=J0041_GMRT_L.ps,width=3.25in,angle=-90}
    \psfig{file=J0041_V8460.ps,width=3.25in,angle=-90}
    }
\caption[]{The GMRT image of J0041+3224 at 1287 MHz with an angular resolution of
$\sim$2.7 arcsec (upper panel) and the VLA C-array image at 8460 MHz with an angular resolution of
$\sim$2.4 arcsec (lower panel).}
\end{figure}

\begin{figure}
\vbox{
    \psfig{file=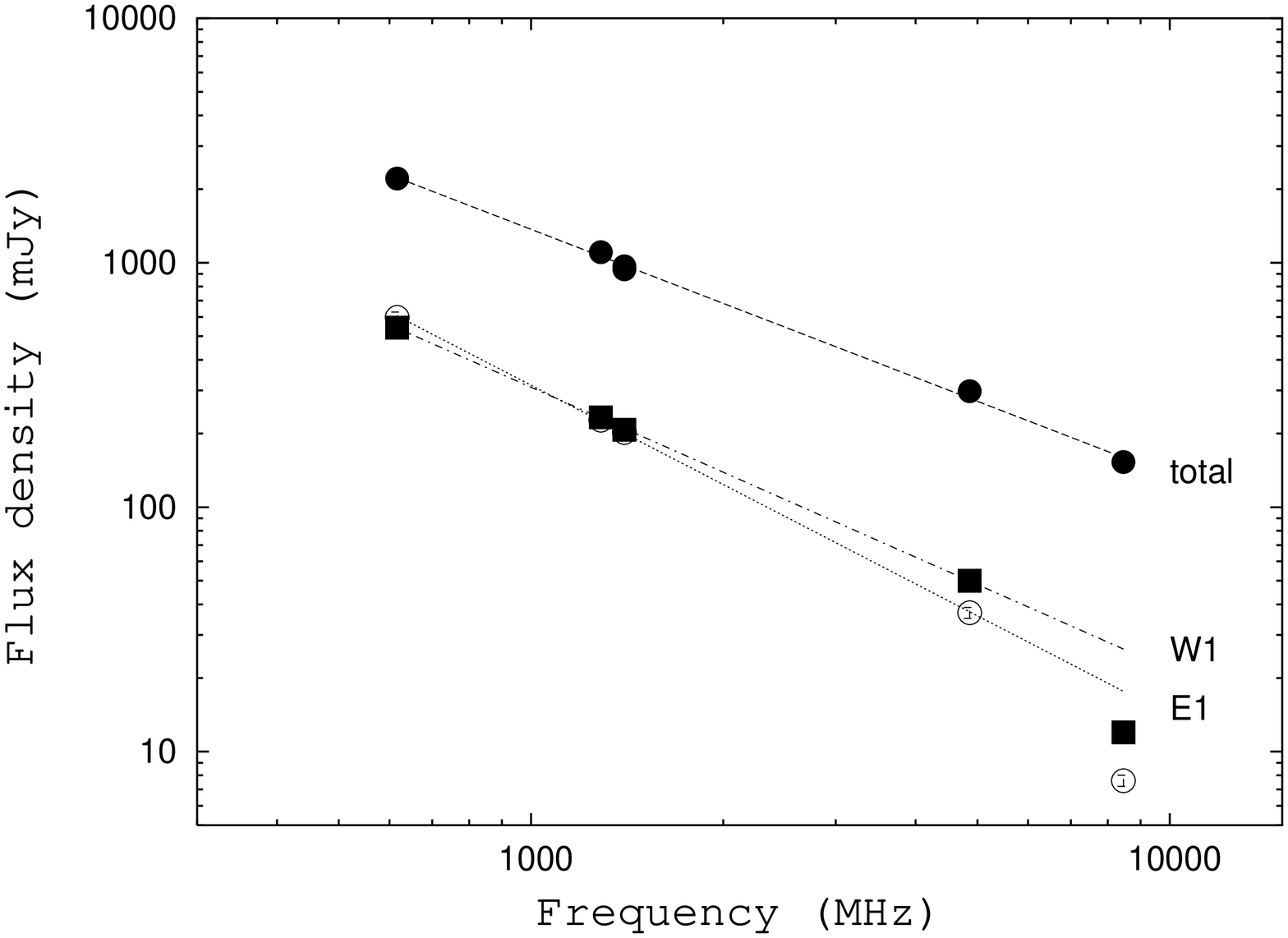,width=3.2in,angle=-90}
    \psfig{file=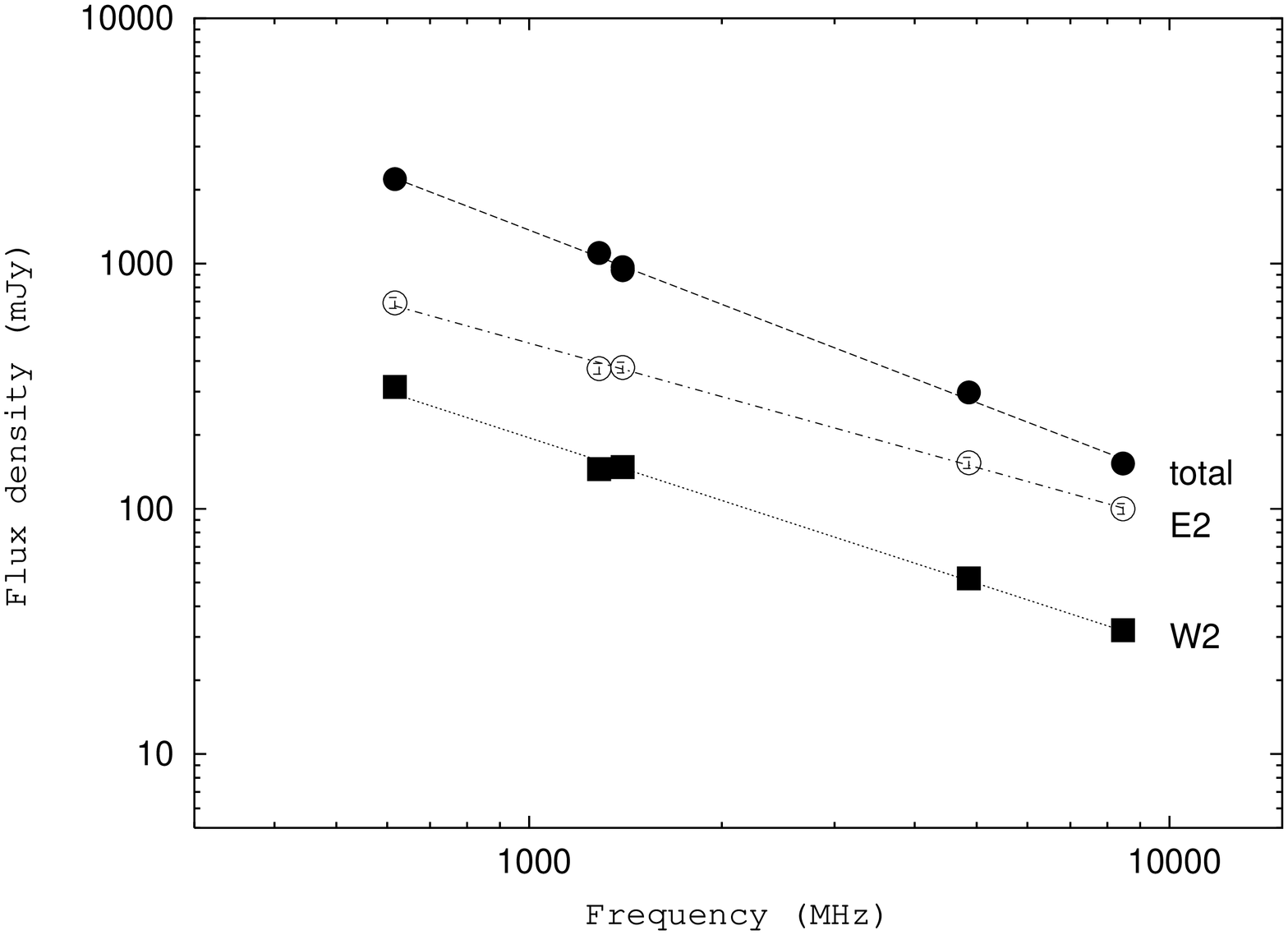,width=3.2in,angle=-90}
    }
\caption[]{The spectra of the western and eastern components of the outer double (upper panel)
and the inner double (lower panel). The integrated spectrum is shown in both the panels.   }
\end{figure}

In Fig. 2 we show the GMRT image at 617 MHz with an angular resolution of $\sim$5.6 arcsec
and the VLA image at 4860 MHz with an angular resolution of $\sim$3.8 arcsec. These images
show the lobes of the inner double to be well separated and lying on opposite sides of
the parent optical galaxy. Weak tails of emission towards the optical galaxy are visible
from both the components of the inner double in the VLA image. These tails merge to form a weak bridge of
emission between the two components.  The spectral index image between 617 and 4860 MHz
obtained by smoothing the VLA image at 4860 MHz to that of the GMRT one is shown superimposed on the 
GMRT image, while the spectral index slices for the western and eastern lobes of the outer
double are shown in Fig. 3. The spectral index image has been made for regions which are at 
least 5 times the rms value in both images. The spectral index, $\alpha$, defined as 
S$\propto\nu^{\alpha}$, varies from $-$0.8 
to $-$1.5 in the western lobe, while for the eastern lobe it varies from $-$1.2 
to $-$1.4. The typical error in the spectral index is $\sim$0.1 assuming an error of
5 per cent in the measured flux density. Using the formalism of 
Myers \& Spangler (1985), the steepening in the western lobe (W1), which has an equipartition
magnetic field (e.g. Miley 1980) of 0.75 nT, corresponds to a spectral 
age $\gapp$6$\times$10$^6$ yr. The eastern component (E1) has a steeper spectrum
but its spectral variation is small.  

\begin{table*}
\caption{ The sample of DDRGs}
                                                                                                                          
\begin{tabular}{l  c  l  rr  c rr rr rr rr l }
                                                                                                                          
\hline
Source       & Opt.& Red- & $l_{in}$& $l_{o}$& Cmp.& R$_{\theta(in)}$ &R$_{\theta(o)}$ & R$_{s(in)}$ &R$_{s(o)}$
& $\Delta_{in}$ & $\Delta_{o}$ & P$_{in}$ & P$_{o}$ & Ref.\\
             & Id. & shift& kpc  & kpc   &     &       &       &        &        & $^\circ$ & $^\circ$ & W/Hz & W/Hz &  \\
    (1)      & (2) &  (3) &  (4) &  (5)  & (6) &  (7)  &  (8)  &  (9)   &  (10)  & (11) & (12)  & (13) & (14)  & (15)  \\
\hline
J0041+3224   &  G  &0.45  & 171  & 969   &W/E  &1.17   &1.49   & 0.39   &1.02    &  0   &  3    & 26.55& 26.52 &1   \\
J0116$-$4722 &  G  &0.146 & 460  &1447   &N/S  &1.06   &1.42   &        &        &  4   & 20    & 25.14& 26.16 &2   \\
J0921+4538   &  G  &0.174 &  69  & 433   &S/N  &3.61   &0.96   &29.00   &1.03    &  1   & 11    & 24.86& 26.82 &3,4,5\\
J0929+4146   &  G  &0.365 & 652  &1875   &N/S  &1.55   &1.06   & 1.12   &0.77    &  3   &  5    & 25.42& 25.64 &6   \\
J1006+3454   &  G  &0.101 & 1.7  &4249   &W/E  &3.15   &0.61   & 0.38   &1.35    & 15   &  2    & 25.86& 25.63 &7,8,9\\
J1158+2621   &  G  &0.112 & 138  & 483   &N/S  &1.23   &0.99   & 1.74   &1.13    &  0   &  1    & 25.31& 25.48 &10  \\
J1242+3838   &  G  &0.300 & 251  & 602   &N/S  &1.74   &0.90   & 2.00   &1.59    &  1   &  3    & 24.32& 24.84 &6   \\
J1247+6723   &  G  &0.107 &0.014 &1195   &     &       &       &        &        &      &       & 24.87& 24.55 &11,12  \\
J1453+3308   &  G  &0.249 & 159  &1297   &S/N  &1.06   &1.34   & 0.16   &0.49    &  3   & 11    & 24.77& 25.90 &6   \\
J1548$-$3216 &  G  &0.108 & 313  & 961   &S/N  &1.24   &1.07   &        &        &  4   &  4    & 24.34& 25.69 &13  \\
J1835+6204   &  G  &0.519 & 369  &1379   &N/S  &1.03   &1.02   & 0.89   &1.89    &  1   &  0    & 26.27& 26.81 &6   \\
J2223$-$0206 &  G  &0.056 & 130  & 612   &S/N  &1.77   &0.95   & 2.66   &1.51    &  2   &  6    & 23.61& 25.58 &14,15 \\
\hline
\end{tabular}
                                                                                                                          
1: Present paper; 2: Saripalli, Subrahmanyan \& Udaya Shankar 2002; 3: Perley et al. 1980;
4: Bridle, Perley \& Henriksen 1986; 5: Clarke et al. 1992;  6: Schoenmakers et al. 2000;
7: Willis, Strom \& Wilson 1974;   8: Strom \& Willis 1980;   9: Schilizzi et al. 2001; 10: Owen \& Ledlow 1997;
11: Marecki et al. 2003;   12: Bondi et al. 2004;  13: Saripalli, Subrahmanyan \& Udaya Shankar 2003;
14: Kronberg, Wielebinski \& Graham 1986; 15: Leahy et al. 1997.
\end{table*}

Our higher-resolution images with the GMRT at 1287 MHz and with the VLA at 8460 MHz 
with angular resolutions of  $\sim$2.7 and 2.4 arcsec respectively are shown in Fig. 4. 
The inner components are resolved with angular
sizes of 1.8$\times$1.0 arcsec$^2$ along a PA of 105$^\circ$ for the western component
and 2.8$\times$0.8 arcsec$^2$ along a PA of 74$^\circ$ for the eastern one. These have
been estimated from the lower-frequency GMRT image. The values are similar using the VLA 
8460-MHz image of similar resolution. The western component of the inner double is extended towards the 
direction of the optical object while the eastern component is extended at about 34$^\circ$
to the axis of the source. This is not surprising given the complexity of the structures
of hotspots and emission in their vicinity (cf. Black et al. 1992; Leahy et al. 1997; 
Hardcastle et al. 1997).

There is a possible detection of a radio core in our data. The peak of emission
in the weak bridge seen in VLA image at 4860 MHz, has a peak flux density of 0.57 mJy/beam
and is within $\sim$1 arcsec of the optical position. If this is indeed the radio core, 
its flux density is likely to be lower since this value could be contaminated by emission
from the bridge.  The peak flux density near the optical position in the 8460-MHz image
is 0.2 mJy/beam and is at the same location as the peak of emission in the 4860-MHz image.
This feature could be the radio core, although it requires confirmation. 

The integrated spectra of the entire source and of the individual components along with
the linear least-square fits to the spectra using our measurements are shown in Fig. 5.
The spectra of the components for the outer and inner doubles are shown in the upper and
lower panels respectively, while the integrated spectrum is shown in both the panels.
The spectral index of the entire source is $-$1.01$\pm$0.02, while that of the western and 
eastern components of the outer double are $-$1.15$\pm$0.01 and $-$1.35$\pm$0.01 respectively. 
The corresponding values for the inner double are $-$0.85$\pm$0.03 and $-$0.73$\pm$0.02 respectively.
For the outer double the fits have been made by excluding the flux densities at 8460 MHz. 
There is some evidence of spectral steepening above 4860 MHz, but we need to make  
lower resolution images at higher frequencies to estimate any missing flux density, and 
hence the degree of spectral steepening. 
  
\section{Discussion and results}
The images of J0041+3224 suggest that there have been two main episodes of activity, 
represented by the outer and inner double-lobed structures which are well aligned and
roughly collinear with the parent optical galaxy. The lobes of the outer double are
separated by 169 arcsec, corresponding to  969 kpc, while the inner double has a separation
of 29.9 arcsec, corresponding to 171 kpc (H$_\circ$=71 km s$^{-1}$ Mpc$^{-1}$, $\Omega_m$=0.27, 
$\Omega_\Lambda$=0.73, Spergel et al. 2003). The radio luminosities of the outer and inner 
doubles at an emitted frequency of 1.4 GHz are 3.3 and 3.6$\times$10$^{26}$ W Hz$^{-1}$ 
respectively, both well above the FRI-FRII divide, and consistent with their observed structures.
 
The lobes of the outer double represent an earlier period of activity. 
These lobes have steeper spectral indices than the
inner lobes (Fig. 5), possible evidence of spectral steepening and no  prominent hotspots, 
while the inner lobes which represent
ongoing activity are dominated by bright hotspots. The ratio of the average peak brightness 
of the inner double to the outer double in our VLA 8460-MHz images which have
angular resolutions of $\sim$2.4 arcsec is $\gapp$40. The angular resolution in these images 
corresponds to a linear size of $\sim$14 kpc, which is within the range for sizes of hotspots
for sources of similar linear size (cf. Leahy et al. 1997; Jeyakumar \& Saikia 2000).

When the jet is interrupted and energy supply to the outer hotspots ceases, the channel will
collapse due to the loss of pressure which is provided by the jet material (e.g. Kaiser \&
Alexander 1997). This implies that the restarted jet will have to drill out a new channel.
However, unlike the earlier jet which propagates through the interstellar and intracluster
or intergalactic medium, the restarted jet propagates into the relic synchrotron plasma
created by the earlier cycle of activity after ploughing its way through the interstellar
medium of the host galaxy. The density of the synchrotron plasma is expected
to be significantly smaller than the intergalactic medium (Clarke \& Burns 1991; Cioffi \& 
Blondin 1992; Loken et al. 1992), by up to a factor of 100, unless there is significant
entrainment (Kaiser et al. 2000). Without entrainment, only weak shocks are expected as
the jet propagates through the synchrotron plasma. In this case, the jets move ballistically
without generating prominent hotspots and cocoons. This appears to be the case, for example,
in J1548$-$3216 (Saripalli et al. 2003). In the case of J0041+3224, the detection of 
prominent hotspots in the inner lobes suggest the existence of significant thermal material,
as has been suggested by Kaiser et al. (2000) to explain the
formation of hotspots in their sample of DDRGs.

An estimate of the time scale of interruption and restarting of jet activity
could provide insights in determining the cause of the interruption. A weak constraint
on the upper limit to the time scale of interruption can be provided by the fact that the
outer lobes are still visible. For a sample of `relic' radio sources, Komissarov \& Gubanov (1994)
have estimated the time scales for the source to fade away to be a few times 10$^7$ yr, which
is comparable to the kinematic age of the source itself. For example, for a velocity of 
advancement of $\sim$0.1c, the kinematic age of the outer double is $\sim$3.2$\times$10$^7$ yr. 
One might expect the jets of the inner double to have higher velocities of advancement if
it encounters only the low-density synchrotron plasma (e.g. Clarke \& Burns 1991). However, 
the formation of strong hotspots suggests it is encountering a dense medium, possibly due
to entrainment, and we assume the velocities to be similar. For a projected linear size of
171 kpc for the inner double, the age of the inner double is 5.6$\times$10$^6$ yr. This leaves
us with a time scale of $\sim$20 Myr between stopping and restarting of the jet. 

\begin{figure}
\hspace{1cm}
\vbox{
    \psfig{file=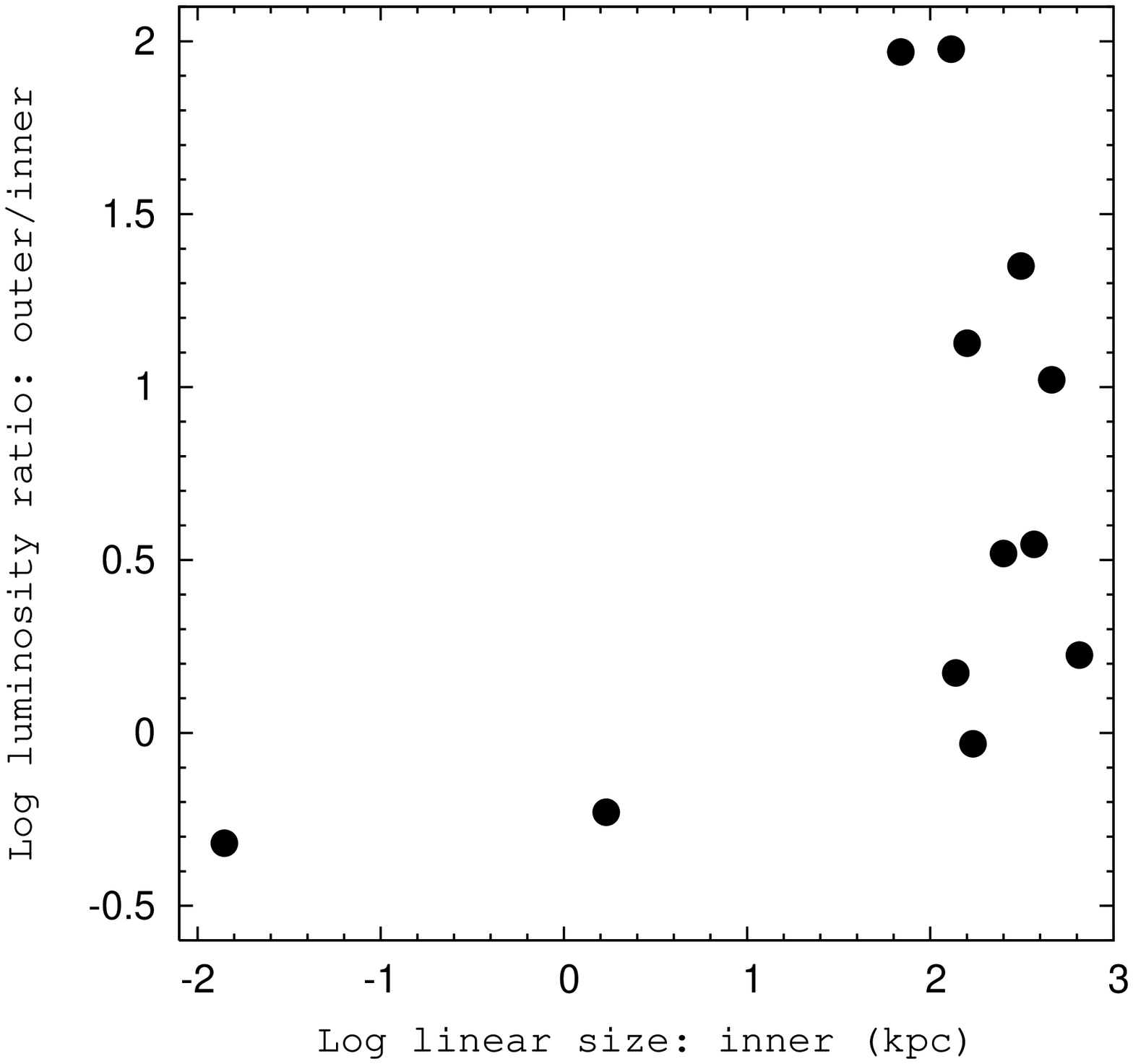,width=3.0in,angle=-90}
    \psfig{file=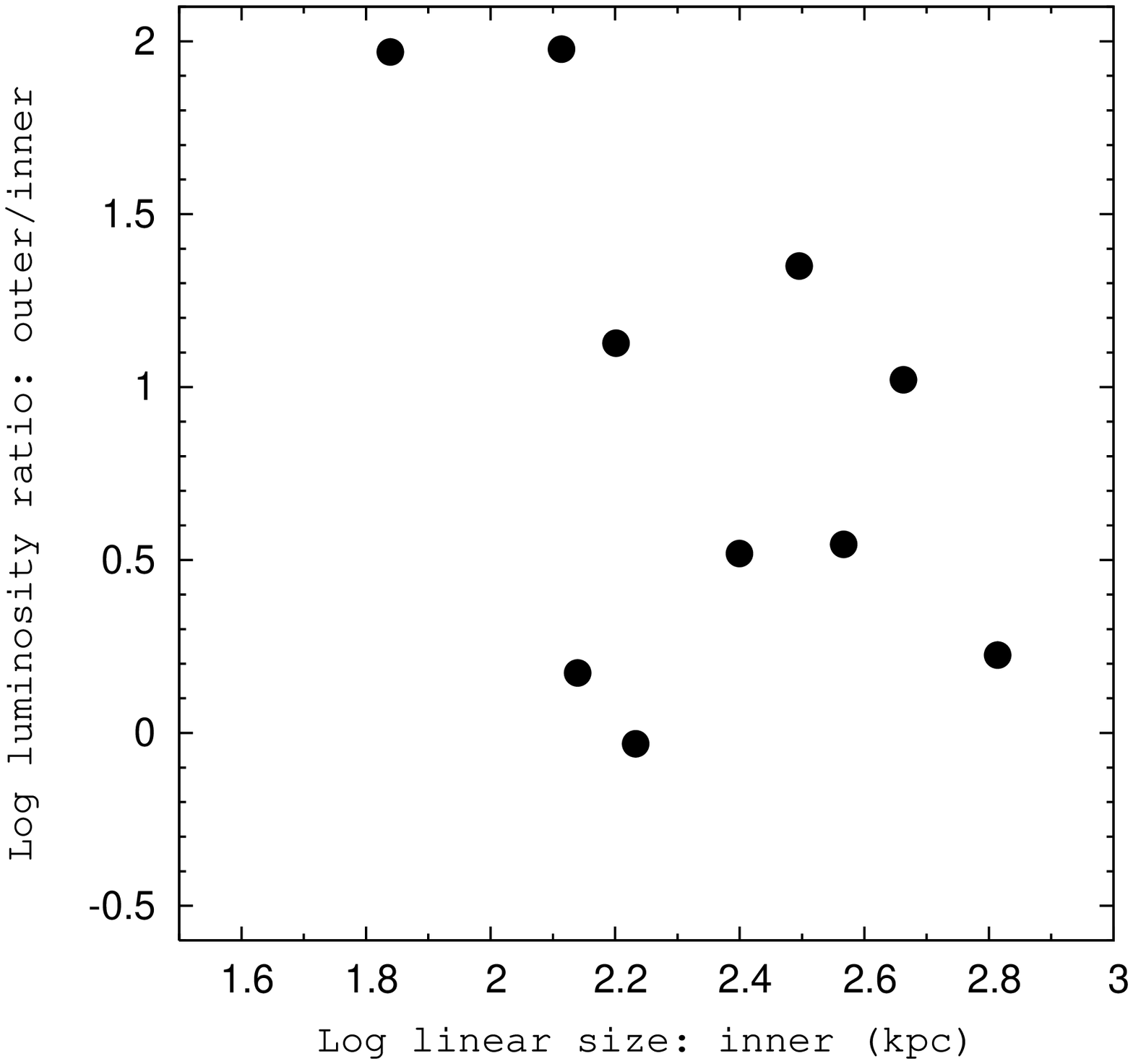,width=3.0in,angle=-90}
    }
\caption[]{Luminosity ratio of the outer double to that of the inner double, P$_{o:in}$, at
an emitted frequency of 1400 MHz is plotted against $l_{in}$, the projected linear size of
the inner double (upper panel). The same plot for those with $l_{in}\gapp$50  kpc is shown
in the lower panel.}
\end{figure}

\subsection{Luminosities of outer and inner doubles}
The luminosities of the outer and inner doubles of J0041+3224 are comparable, the ratio,
P$_{o:in}$, being $\sim$0.93 at an emitted frequency of 1400 MHz. This is unlike the sample of DDRGs
compiled by S2000 where the outer lobes were always found to be 
more luminous than the inner ones. S2000 also noted that P$_{o:in}$
decreases with, $l_{in}$, the separation of the
inner double. We have re-examined this relationship by enlarging the sample of DDRGs
to include those which have been reported more recently, as well as those where the 
inner doubles are of subgalactic dimesnions, as in J1006+3454 (3C236) and J1247+6723. 
Candidate DDRGs mentioned by S2000 such as 4C12.03, 3C16 (Leahy \& Perley 1991) and 
3C424 (Black et al. 1992) which require further observations to clarify their structures
have not been presently included in the sample. The sample is listed in Table 3 and 
is arranged as follows. Column 1: source name; column 2: optical identification;
column 3: redshift; columns 4 and 5:  projected linear size of the inner and outer
double-lobed source in kpc; columns 6: the locations of the components farther/closer
from the core for the inner double. The symmetry
parameters in columns 7 to 10 are all in the same sense as in column 6.  Columns 7
and 8: the arm-length or separation ratio for the inner and outer doubles;
columns 9 and 10: flux density ratios for the inner and outer doubles; 
columns 11 and 12: the misalignment angles, defined to be the supplement of the angle
formed at the core by the hotspots or peaks of emission for the inner and outer lobes;
columns 13 and 14: log of radio luminosity at an emitted frequency of 1.4 GHz for the inner
and outer doubles; column 15: references for the radio structure.

\begin{figure}
\vbox{
    \psfig{file=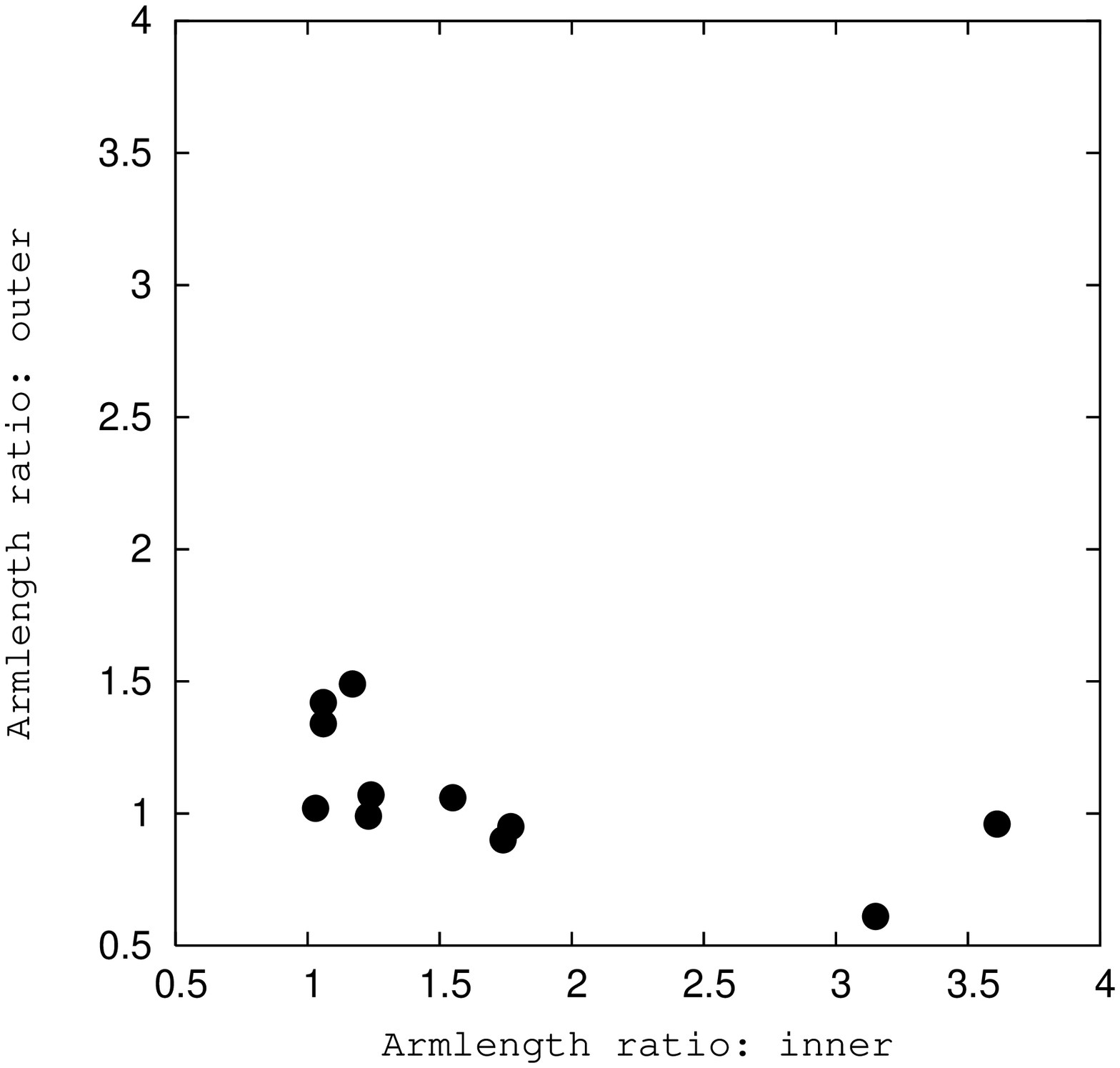,width=3in,angle=-90}
    \psfig{file=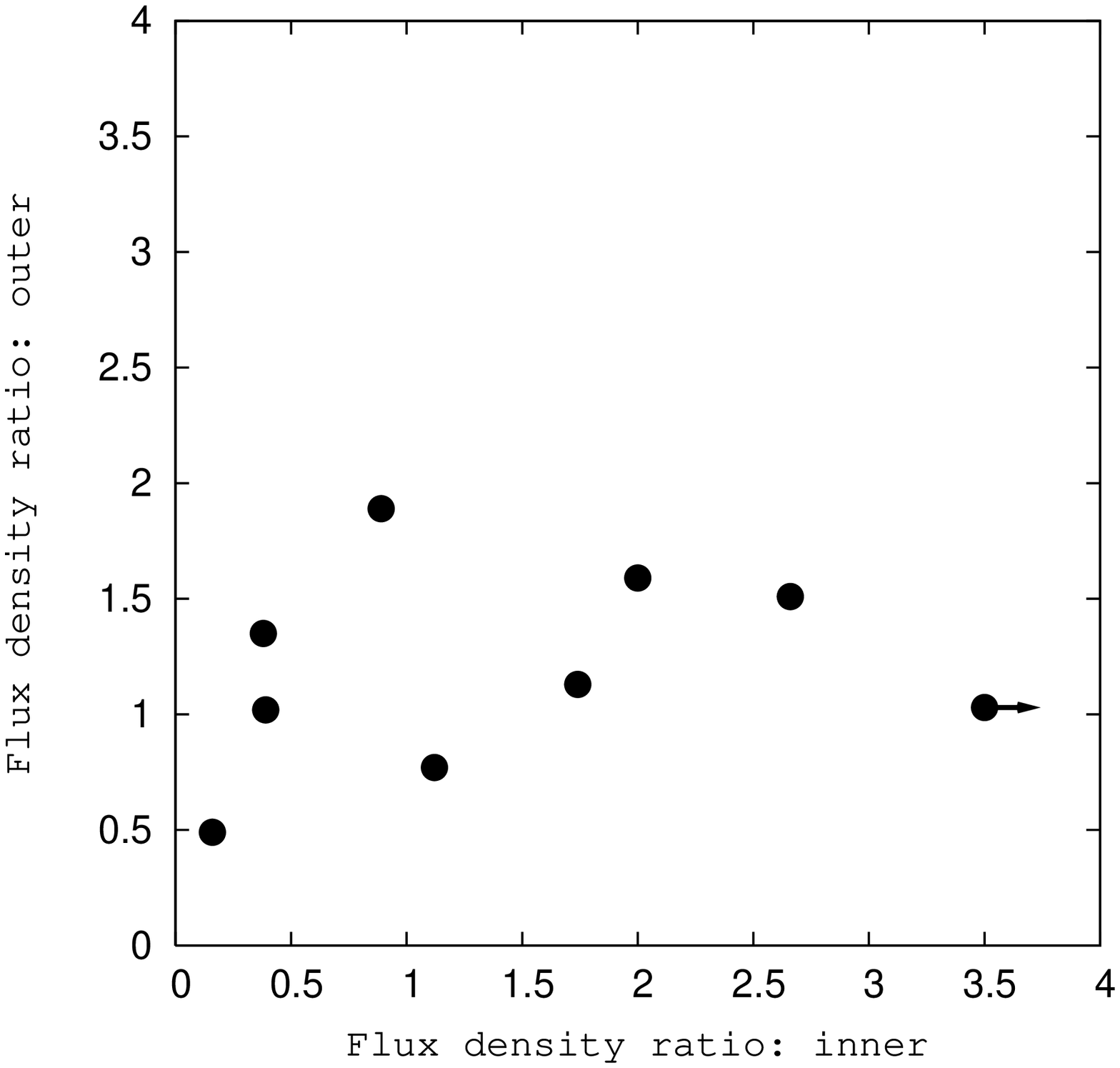,width=3in,angle=-90}
    }
\caption[]{The armlength (upper panel) and flux density (lower panel) ratios of the inner
doubles are plotted against the corresponding values for the outer doubles for the sample of DDRGs.
The sense of direction of the components is described in Section 4.1 and listed in Table 3. }
\end{figure}

In Fig. 6, we plot the ratio of the luminosities of the outer to the inner doubles
against the separation of the inner double. Both the sources with $l_{in} \lapp$1 kpc
have a more luminous inner double than the outer one, unlike all the sources in the
sample of S2000 where $l_{in}\gapp$50 kpc. For comparison with the
plot by S2000  we have also plotted the sources with $l_{in}\gapp$50 kpc separately.  
It is possible that in the early phase of
the evolution of the inner double, where it is ploughing its way through the dense
interstellar medium, conversion of beam energy into radio emission may be more efficient.
This could lead to a ratio of the luminosities of the outer to the inner doubles being
significantly less than unity. As the source expands and traverses through a more
tenuous medium, the ratio could increase with size before approaching values of unity
for large values of $l_{in}$ (cf. S2000). We have re-examined the inverse correlation
suggested by S2000 considering only those objects with $l_{in}\gapp$50 kpc
and find the correlation to have a Spearman rank correlation coefficient of $-$0.37,
compared with a value of $-$0.57 for the objects in the sample of S2000. 

\subsection{Symmetry parameters}
A comparison of the symmetry parameters of the inner and outer doubles might provide
insights into the environments in which the jets are propagating as well as any
possible intrinsic asymmetries in the jets themselves. In the case of the restarted
jets advancing into synchrotron relics from earlier epochs of activity, one might
expect similar environments and hence more symmetric structures for the inner doubles.

\begin{figure}
\vbox{
    \psfig{file=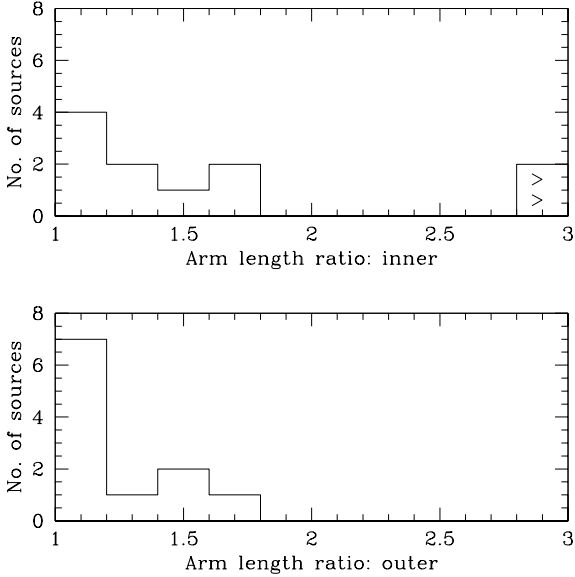,width=3.2in,angle=0}
    }
\caption[]{The distributions of the armlength ratios for the inner and outer doubles, defined
to be $\geq$1. The $>$ sign indicates a value larger than the range plotted in the histogram. }
\end{figure}

\begin{figure}
\vbox{
    \psfig{file=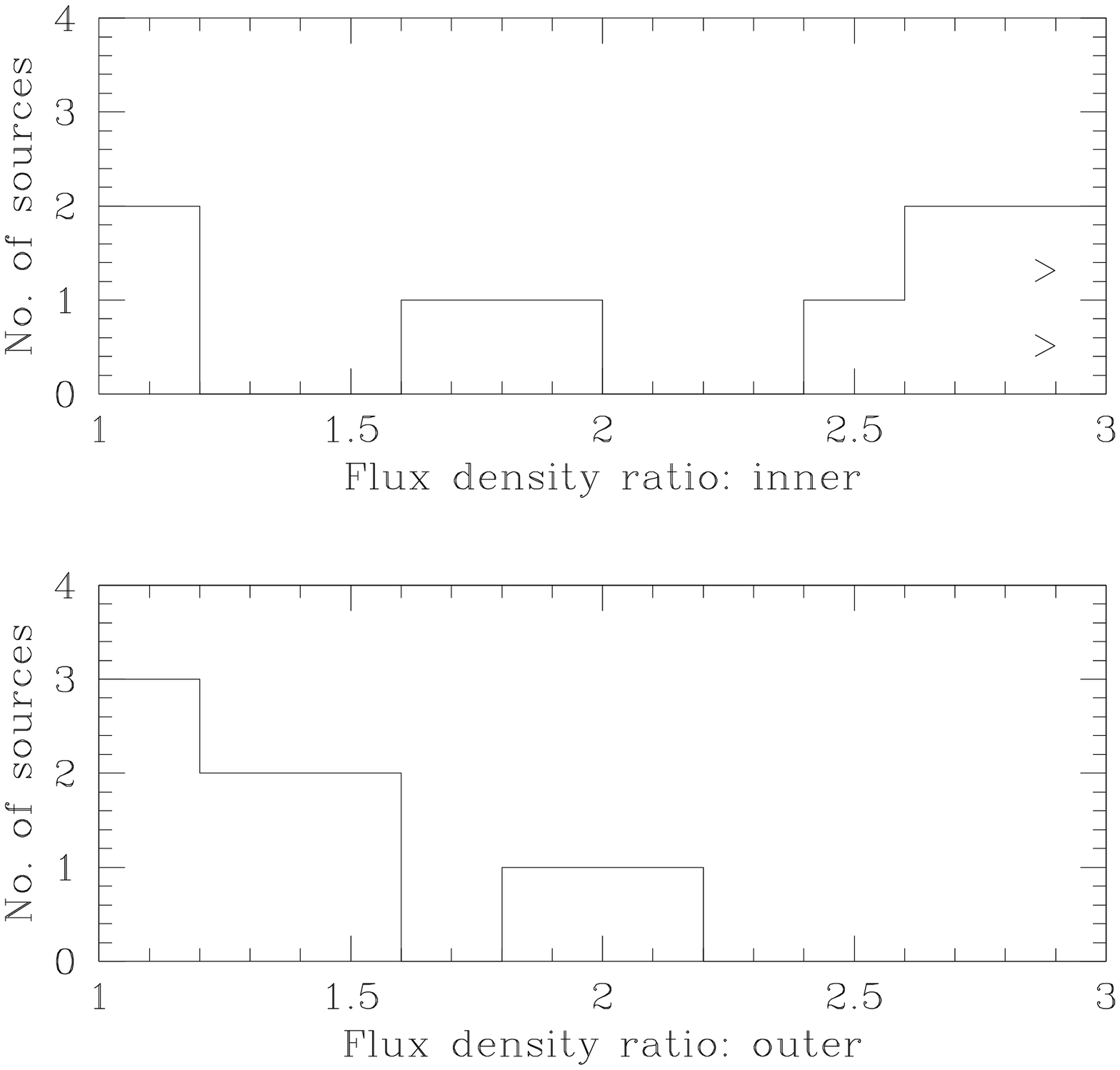,width=3.2in,angle=0}
    }
\caption[]{The distributions of the flux density ratios for the inner and outer doubles, defined
to be $\geq$1. The $>$ sign indicates a value larger than the range plotted in the histogram. }
\end{figure}

\begin{figure}
\vbox{
    \psfig{file=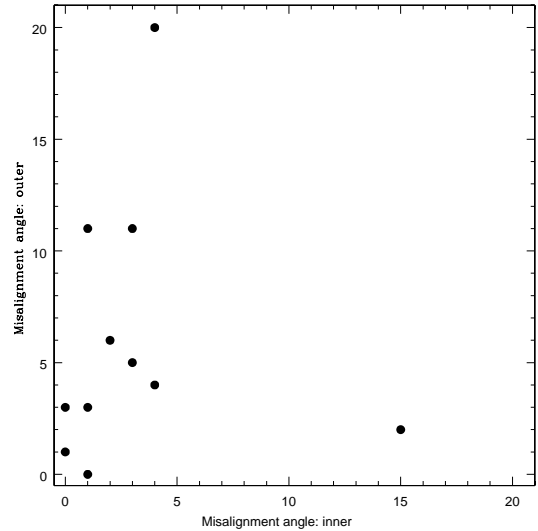,width=3.0in,angle=0}
    }
\caption[]{The misalignment angle of the outer double in degrees is plotted against the corresponding value
for the inner double.}
\end{figure}

In Fig. 7, we show the armlength and flux density ratios of the outer doubles plotted against
the corresponding values for the inner doubles for the sample of DDRGs. The sense of
direction of the components is described in Section 4.1 and listed in Table 3.
It appears that the inner doubles tend to
be more asymmetric in both its armlength and flux density ratios compared with the outer
doubles. The distributions for the armlength ratios, now defined to be always $>$1,
show that the median values are
$\sim$1.3 and 1.1 for the inner and outer doubles respectively (Fig. 8). A Kolmogorov--Smirnov test
shows the distributions to be different at a significance level of 0.30. A
similar trend is also seen in the distributions of the flux density ratios, again defined to
be always $>$1, for the inner
and outer lobes. The median values are $\sim$2.5 and 1.4 respectively (Fig. 9), and the two distributions
are different at a significance level of 0.10. It is also worth noting (see Table 3) that
the asymmetries in the inner and outer doubles are not in the same sense. Although these
trends need to be confirmed with larger samples, they are possibly a reflection of
different environments, possibly due to different degrees of entrainment in the cocoons on 
opposite sides coupled with effects of relativistic beaming and any intrinsic jet asymmetries.
For example, in the case of J0921+4538, the armlength ratio of 3.61 for the inner double can be understood
in the relativistic beaming framework if the velocity is $\sim$0.8c for an inclination angle of 
45$^\circ$, the dividing line between radio galaxies and quasars in the unified scheme (Barthel 1989).
This is also in reasonable agreement with the flux density ratio of 29. However, in the case of the inner
double of  J1006+3454, where the armlength ratio is 3.15, the flux density ratio is 0.38, suggesting
intrinsic asymmetries.

The median values of the misalignment angle of the inner and outer doubles  are 
$\sim$2 and 4$^\circ$ respectively (Fig. 10). A Kolmogorov--Smirnov test shows the distributions to be 
different at a significance level of 0.3.  This could arise due to the
lobes of the outer double responding to either large-scale density gradients 
or motion of the parent optical galaxy during the two cycles of jet activity.

\section{Concluding remarks}
We have reported the discovery of a new double-double radio galaxy, J0041+3224, with the
GMRT and subsequent observations with the VLA. Using an estimated redshift of 0.45, the 
projected linear size of the outer double is 969 kpc. Large linear sizes are characteristic
of most of the known DDRGs. The lobes of the outer double have steeper spectral indices 
compared with those of the inner double. The kinematic age of the outer double is 
$\sim$3$\times$10$^7$ yr, while the time scale of interruption of jet activity is
$\sim$20 Myr.  Unlike most DDRGs with $l_{in}$$\gapp$50 kpc, the inner double 
of J0041+3224 is marginally more luminous than the outer one. S2000 reported an
inverse correlation between the ratio of the luminosities of the outer and inner 
doubles, P$_{o:in}$, and $l_{in}$. For their sample of DDRGs 
the luminosity ratio, P$_{o:in}$, is as high as $\sim$100 for $l_{in}$$\sim$50 kpc and approaches
unity when $l_{in}$ approaches values close to a Mpc, the ratio being always $>$1. Considering 
the more compact inner doubles with $l_{in}$ $\lapp$ few kpc, namely J1006+3454 (3C236) and 
J1247+6723, the inner doubles are significantly more luminous than the outer ones. This 
suggests that in the early phase of
the evolution of the inner double, where it is ploughing its way through the dense
interstellar medium, conversion of beam energy into radio emission may be more efficient.
In this case, the ratio of the luminosity of the outer to the inner double could increase
with size before decreasing and approaching a value of about unity when $l_{in}$ 
approaches $\sim$1 Mpc.
We have re-examined the inverse correlation for sources with $l_{in}$$\gapp$50 kpc 
with the addition of a few more sources and find the correlation to have a rank correlation
co-efficient of $-$0.37, compared with a value of $-$0.57 for the objects in the sample of S2000.
   
We have compared the symmetry parameters of the inner and outer doubles and find that the
inner doubles appear to be more asymmetric in both its armlength and flux density ratios 
compared with the outer doubles. Also, the asymmetries in the inner and outer doubles are 
not in the same sense. Although these trends need to be confirmed with larger samples, 
they are possibly a reflection of different environments due to different degrees of 
entrainment in the cocoons on opposite sides, coupled with any intrinsic jet asymmetries
and effects of relativistic motion. However, the inner doubles appear marginally 
more collinear with the radio core than the outer doubles. 
This could arise due to the lobes of the outer double responding to large-scale density gradients, 
or motion of the parent optical galaxy during the two cycles of nuclear activity.

\section*{Acknowledgments} 
We thank our colleagues for useful discussions, an anonymous referee, Dave Green  and Paul Wiita
for their comments on the manuscript  and the telescope operators for their help with the observations.
The Giant Metrewave Radio Telescope is a national facility operated by the National Centre 
for Radio Astrophysics of the Tata Institute of Fundamental Research.
The National Radio Astronomy Observatory  is a
facility of the National Science Foundation operated under co-operative
agreement by Associated Universities Inc. 
This research has made use of the NASA/IPAC extragalactic database (NED)
which is operated by the Jet Propulsion Laboratory, Caltech, under contract
with the National Aeronautics and Space Administration.

{}


\begin{thebibliography}{}

\bibitem[]{}  Baars J.W.M., Genzel R., Pauliny-Toth I.I.K., Witzel A. 1977, A\&A, 61, 99
\bibitem[]{}  Barthel P.D., 1989, ApJ, 336, 606
\bibitem[]{}  Baum S.A., O'Dea C.P., de Bruyn A.G., Murphy D.W., 1990, A\&A, 232, 19
\bibitem[]{}  Black A.R.S., Baum S.A., Leahy J.P., Perley R.A., Riley J.M., Scheuer P.A.G., 1992, MNRAS, 256, 186
\bibitem[]{}  Bondi M., March\~a, M.J.M., Polatidis A., Dallacasa D., Stanghellini C., Ant\'on S., 2004, MNRAS, 352, 112
\bibitem[]{}  Bridle A.H., Perley R.A., Henriksen R.N., 1986, AJ, 92, 534
\bibitem[]{}  Burns J.O., Feigelson E.D., Schreier E.J., 1983, ApJ, 273, 128
\bibitem[]{}  Burns J.O., Schwendeman E., White R.A., 1983, ApJ, 271, 575
\bibitem[]{}  Cioffi D.F., Blondin J.M., 1992, ApJ, 392, 458
\bibitem[]{}  Clarke D.A., Burns J.O., 1991, ApJ, 369, 308
\bibitem[]{}  Clarke D.A., Bridle A.H., Burns J.O., Perley R.A., Norman M.L., 1992, ApJ, 385, 173
\bibitem[]{}  Clarke D.A., Burns J.O., Norman M.L., 1992, ApJ, 395, 444
\bibitem[]{}  Franceschini A., Vercellone S., Fabian A.C., 1998, MNRAS, 297, 817
\bibitem[]{}  Gizani N.A.B., Leahy J.P., 2003, MNRAS, 342, 399
\bibitem[]{}  Guiderdoni B., Rocca-Volmerange B., 1987, A\&A, 186, 1                                      
\bibitem[]{}  Hardcastle M.J., Alexander P., Pooley G.G., Riley J.M., 1997, MNRAS, 288, 859
\bibitem[]{}  Ivezi\'{c} Z., et al. 2002, AJ, 124, 2364
\bibitem[]{}  Ivezi\'{c} Z., et al. 2004, ASPC, 311, 347
\bibitem[]{}  Jeyakumar S., Saikia D.J., 2000, MNRAS, 311, 397
\bibitem[]{}  Junkes N., Haynes R.F., Harnett J.I., Jauncey D.L., 1993, A\&A, 269, 29
\bibitem[]{}  Kaiser C.R., Alexander P., 1997, MNRAS, 286, 215
\bibitem[]{}  Kaiser C.R., Schoenmakers A.P., R\"{o}ttgering H.J.A., 2000, MNRAS, 315, 381
\bibitem[]{}  Komissarov S.S., Gubanov A.G., 1994, A\&A, 285, 27
\bibitem[]{}  Kronberg P.P., Wielebinski R., Graham D.A. 1986, A\&A, 169, 63
\bibitem[]{}  Leahy J.P., Perley R.A., 1991, AJ, 102, 537
\bibitem[]{}  Leahy J.P., Pooley G.G., Riley J.M., 1986, MNRAS, 222, 753                
\bibitem[]{}  Leahy J.P., Black A.R.S., Dennett-Thorpe J., Hardcastle M.J., Komissarov S., Perley R.A., 
              Riley J.M., Scheuer P.A.G., 1997, MNRAS, 291, 20
\bibitem[]{}  Loken C., Burns J.O., Clarke D.A., Norman M.L., 1992, ApJ, 392, 54
\bibitem[]{}  Marconi A., Risaliti G., Gilli R., Hunt L.K., Maiolino R., Salvati M., 2004, MNRAS, 351, 169
\bibitem[]{}  Marecki A., Barthel P.D., Polatidis A., Owsianik I., 2003, PASA, 20, 16 
\bibitem[]{}  Miley G. K., 1980, ARA\&A, 18, 165
\bibitem[]{}  Myers S. T., Spangler S. R., 1985, ApJ, 291, 52
\bibitem[]{}  Morganti R., Killeen N.E.B., Ekers R.D., Oosterloo T.A., 1999, MNRAS, 307, 750
\bibitem[]{}  Nipoti C., Blundell K.M., Binney J., 2005, MNRAS, 361, 633
\bibitem[]{}  Owen F.N., Ledlow M.J., 1997, ApJS, 108, 41
\bibitem[]{}  Padrielli L., Kapahi V.K., Katgert-Merkelijn J.K., 1981, A\&AS, 46, 473	
\bibitem[]{}  Perley R.A., Bridle A.H., Willis A.G., Fomalont E.B., 1980, AJ, 85, 499
\bibitem[]{}  Roettiger K., Burns J.O., Clarke D.A., Christiansen W.A., 1994, ApJ, 421, 23L
\bibitem[]{}  Saikia D.J., Thomasson P., Spencer R.E., Mantovani F., Salter C.J., Jeyakumar S., 2002, A\&A, 391, 149
\bibitem[]{}  Saripalli L., Subrahmanyan R., Udaya Shankar N., 2002, ApJ, 565, 256
\bibitem[]{}  Saripalli L., Subrahmanyan R., Udaya Shankar N., 2003, ApJ, 590, 181
\bibitem[]{}  Schilizzi R.T. et al., 2001, A\&A, 368, 398
\bibitem[]{}  Schoenmakers A.P., de Bruyn A.G., R\"{o}ttgering H.J.A., van der Laan H., Kaiser C.R., 2000, MNRAS, 315, 371
\bibitem[]{}  Spergel D.N. et al., 2003, ApJS, 148, 175
\bibitem[]{}  Stanghellini C., O'Dea C.P., Dallacasa D., Cassaro P., Baum S.A., Fanti R., Fanti C., 2005, A\&A, submitted,
              (astro-ph/0507499)
\bibitem[]{}  Strom R.G., Willis A.G., 1980, A\&A, 85, 36
\bibitem[]{}  Subrahmanyan R., Saripalli L., Hunstead R.W., 1996, MNRAS, 279, 257
\bibitem[]{}  van Breugel W., Fomalont E.B., 1984, ApJ, 282, 55L
\bibitem[]{}  Willis A.G., Strom R.G., Wilson A.S., 1974, Nature, 250, 625
 
\end{thebibliography}
\end{document}